\documentstyle[12pt]{article}

\def\la{\mathrel{\mathpalette\fun <}}

\def\fun#1#2{\lower3.6pt\vbox{\baselineskip0pt\lineskip.9pt
\ialign{$\mathsurround=0pt#1\hfil##\hfil$\crcr#2\crcr\sim\crcr}}}

\begin{document}

\title{ Problem of the complete measurement for $
CP-$violating parameters in neutral $B-$meson decays}

\author{Ya.I. Azimov\thanks{azimov@lnpi.spb.su}, V.L.
Rappoport,\ V.V.  Sarantsev\\ Petersburg Nuclear Physics
Institute\\ Gatchina, St.Petersburg 188350, Russia}
\maketitle

\begin{abstract}
Phenomenological $CP$-violating parameters in decays of neutral
$B$-mesons are discussed. Special attention is given to the degree of
their measurability. We emphasize important role of the sign of $\Delta
m_B$ and suggest how it could be determined experimentally.
\end{abstract}

\noindent PACS: 11.30.Er, 14.40.Jz

\newpage
\section {Introduction}

Two opinions may be considered now as generally accepted (see,
e.g., reviews \cite{1,2}):
\begin{enumerate}
\item The origin of $CP-$violation will not be established till its
manifestations are known only for neutral kaons.

\item  The most promising test-ground for detailed studies of
$CP-$violation is given by decays of neutral $B-$mesons.
\end{enumerate}

As a result, much work was devoted to discussion of $B-$meson decay
modes favourable for $CP-$violation searches and to experimental
manifestations of possible sources of the violation (see, e.g.,
references in \cite{1,2}). A more straightforward problem, degree
of measurability of phenomenological parameters describing
$CP-$violation in $B-$meson decays, has not been considered.
One possible reason could be a close similarity of neutral
$B-$mesons to neutral kaons.  But heavier masses of the third
quark generation produce various differences, sometimes rather
essential, in the meson decay properties. Therefore, in the
present paper we reanalyse basic $CP-$violating parameters
for $B-$mesons. Special attention is given to the question how one
could achieve their complete measurement.

The presentation goes as follows. In Section 2 we discuss various
parameters that are familiar to express $CP-$violation. Of special
interest is degree of their rephasing (in)dependence. Section 3
considers how the parameters appear in experimental lifetime
distributions. The role of width and mass differences for
measurability of phenomenological parameters is emphasized.
This consideration is continued in Section 4 where we also
suggest how one could measure the sign of $\Delta m_B$
providing the basis for the complete measurement of all $CP-$violating
parameters. Summary of our results and their short discussion are
given in the last Section.

\section{$CP-$violating parameters}

As is well-known the time evolution of neutral $B-$mesons is
determined by two states
\begin{equation}
B_\pm\ =\ \frac{1}{\sqrt{2(1+|\varepsilon_B|^2)}}\
\left[(1+\varepsilon_B)B^0 \pm(1-\varepsilon_B)\overline B^0\right]\ .
\end{equation} If we
use the phase convention
\begin{equation} \overline B^0\ =\ (CP)B^0\ , \end{equation}
the exact
$CP-$conservation would lead to $\varepsilon_B=0$ and the states
$B_\pm$ having the definite $CP-$parity equal $\pm 1$.
Generally, they are eigen-states of the effective
(non-Hermitian) Hamiltonian. But possibility of rephasing
$\overline B^0$ (with appearance of a phase-factor in relation
(2) but without changing $B_+$ and $B_-$) means that
$\varepsilon_B$ itself can not be measurable. Only $\left
|\frac{1+\varepsilon_B}{1-\varepsilon_B}\right |$ is
rephasing-invariant and admits measurement. The value
\begin{equation}
\delta_B\ =\ \frac{|1+\varepsilon_B|^2-|1-\varepsilon_B|^2}
{|1+\varepsilon_B|^2+|1-\varepsilon_B|^2}\ =\ \frac{2{\rm
Re}\,\varepsilon_B} {1+|\varepsilon_B|^2} \end{equation} may be considered as
the measure of $CP-$violation in  $B\overline B$ mixing.

Similar quantity for neutral kaons, $\delta_K$, describes the
charge asymmetry in semileptonic decays of $K_L$ (or in decays
of pure $K_S$ if it could be separated). The same would be true
for $B-$mesons as well. But the real separation of $B_L$ and
$B_S$ is hardly possible because of expected smallness of
$(\Gamma_S-\Gamma_L)_B$.  Therefore, the problem arises how to
find ways of physical identification for the states $B_\pm$ in
the absence of $CP-$conservation.

Mixing of $B^0$ and $\overline B^0$ is mainly determined by
the very heavy intermediate state $t\overline t$ . The
corresponding transition amplitude contains quarks from two
generations only. Since $CP-$violation (by the standard
CKM-\-mechanism) requires participation of all three quark
generations, the value of $\delta_B$ has an additional
kinematic suppression by the factor $m_c^2/m_t^2$
(more detailed discussion see, e.g., in \cite{3}).
Thus, contrary to neutral kaons, there are no hopes to find
experimentally nonvanishing $\delta_B$ in near future.

More promising are studies of decays
\begin{equation}
B^0(\overline B^0)\ \rightarrow\ f
\end{equation}
with final states $f$ having definite $CP-$parity [3,4]. As the
measure of $CP-$violation for a particular decay one may use
deviation of the parameter
\begin{equation}
\lambda_B^{(f)}\ =\ \frac{1-\varepsilon_B}{1+\varepsilon_B}\ \cdot\
 \frac{\langle f|\overline B^0\rangle}{\langle f|B^0\rangle}
\end{equation}
 from the $CP-$parity value of the state $f$. Any
 $\lambda_B^{(f)}$ is rephasing-invariant and, hence, its
complete measurement (i.e., of both the absolute value and
phase) should be possible.

Now we can restrict ourselves to considering such independent
states where final  state interaction produces only elastic
rescattering (really we mean states that diagonalize
$S-$matrix of strong interactions; analogues in neutral kaon
decays are, e.g., two-\-pion states with definite isospin
values).  Any other final state can be expanded as series over
the independent ones.

Assumption of $CPT-$invariance leads to the conclusion that
the ratio of amplitudes in expression (5) for an independent
final state is a phase factor. So we have
\begin{equation} |\lambda_B^{(f)}|\ =\ \left
|\frac{1-\varepsilon_B}{1+\varepsilon_B}\right | \end{equation} for every
independent state $f$.

Thus, the parameter $\delta_B$ generated by the $CP-$violation
in mixing appears to be universal and determines deviation of
any $|\lambda_B^{(f)}|$ from unity. Only one additional
$CP-$violating parameter, $\arg(\lambda_B^{(f)})$, may arise for
each particular independent final state $f$ in the neutral
$B-$decays. They are just the parameters that phenomenologically
correspond to direct $CP-$violation in the particular decay
mode.

Traditional $CP-$violating parameters $\eta$, similar to ones used
for kaon decays, are simply related to $\lambda$'s. For $CP-$even
states $f+$~:
\begin{equation}
\eta_B^{(f+)}\ =\ \frac{1-\lambda_B^{(f+)}}{1+\lambda_B^{(f+)}}\ ;
\end{equation}
for $CP-$odd state $f-$~:
\begin{equation}
\eta_B^{(f-)}=\frac{1+\lambda_B^{(f-)}}{1-\lambda_B^{(f-)}}\ .
\end{equation}
Comparing eqs.(5),(7),(8) shows that one is always able to find
an appropriate phase convention for $B$ and $\overline B$ which
changes  $\varepsilon_B$ so to equate
\begin{equation}
\varepsilon_B\ =\ \eta_B^{(f)}
\end{equation}
for any chosen independent state $f$. Independently of any phase
convention we have
\begin{equation}
\delta_B\ =\ \frac{2{\rm Re}\,\eta_B^{(f)}}{1+|\eta_B^{(f)}|^2}\ .
\end{equation}

Note that all the above relations, including eq.(10), are true for
neutral kaons as well. This leads to new experimental predictions
based on the $CPT-$invariance. E.g., $CP-$violating parameters
in decays $K^0(\overline K^0)\rightarrow 2\pi$ and
$K^0(\overline K^0)\rightarrow 3\pi$ should satisfy equality
\begin{equation}
\frac{{\rm Re}\,\eta_K^{(2\pi)}}{1+|\eta_K^{(2\pi)}|^2}\ =\
\frac{{\rm Re}\,\eta_K^{(3\pi)}}{1+|\eta_K^{(3\pi)}|^2}\ =\ \frac
12\ \delta_K
\end{equation}
(we assume here both $2\pi$ and $3\pi$ states to have a definite
isotopic structure). Thus, measurement of the corresponding
$|\eta|^2$ immediately leads to determination of $\arg\eta$
(double-valued, up to the sign of ${\rm Im}\eta$). Eq.(11) should be
applicable also to the decay $K^0(\overline K^0)\rightarrow\pi^+
\pi^-\gamma$ where $CP-$violation has been observed experimentally
\cite{5}. It does really work within available precision.

Specific feature of neutral $B-$mesons, having no analogues for
neutral kaons, is the presence of decays \begin{equation} B^0(\overline
B^0)\ \rightarrow\ fK^0(\overline K^0)\ , \end{equation} with $f$, again,
being definite $CP-$parity states. They are induced by the quark
decay $b\rightarrow c\overline c s$. The most popular final
state of such a kind is $J/\psi K^0(\overline K^0)$. Unique
property of decays (12) is the coherence of neutral $B$ and neutral
$K$ evolutions \cite{6}.

Decays (12) generate new set of parameters:
\begin{equation}
\lambda_{BK}^{(f)}\ =\ \frac{1-\varepsilon_B}{1+\varepsilon_B}\ \cdot\
\frac{1+\varepsilon_K}{1-\varepsilon_K}\ \cdot\ \frac{\langle f\overline
K^0| \overline B^0\rangle}{\langle fK^0|B^0\rangle}
\end{equation} similar to
(5).  They are invariant under rephasing of both $B$ and $K$ mesons
and should also be completely measurable. If the final system
may be considered as independent (in the above sense) then the
ratio of amplitudes in eq.(13) is again a phase factor and
deviation of $|\lambda_{BK}^{(f)}|$ from unity becomes
universal. But it is influenced, differently from
$|\lambda_B^{(f)}|$, by the $CP-$violation in both $B-$ and
$K-$mixing.

\section{Experimental manifestation and \mbox{measurability}}

To suggest ways for the complete measurement of the parameters
$\lambda$ we should first consider how they reveal
themselves experimentally. The problem of measurement for
$\delta_B$ looks quite clear and we will not discuss it here anymore.
Situation is not so clear for parameters $\lambda$.

Standard calculations for the
decay (4) in the case of the initially pure $B^0-$meson lead to the
time distribution \begin{equation} W_B^{(f)}(t)\ \sim\
\left|\frac{1+\lambda_B^{(f)}}{2}\right|^2
\exp(-\Gamma_+t)+\left|\frac{1-\lambda_B^{(f)}}{2}\right|^2
\exp(-\Gamma_-t) \ +
\end{equation}
$$+\ \exp\left(-\ \frac{\Gamma_+ +\Gamma_-}{2}\ t\right)\left(
\frac{1-|\lambda_B^{(f)}|^2} {2}\cos\Delta m_Bt-{\rm
Im}\lambda_B^{(f)}\cdot\sin\Delta m_Bt\right)\ .$$
Here $\Delta m_B=m_--m_+$; $m_+, \Gamma_+$ and
$m_-, \Gamma_-$ are the mass and width
of the corresponding state $B_+$ or $B_-$.
To obtain the distribution for the initially pure
$\overline B^0-$meson one should change
$\lambda\rightarrow 1/\lambda$. Eq.(14) has the same
structure as, e.g., distribution of decays
$K^0(t)\rightarrow\pi\pi$. The first two terms are contributions
of states $B_{\pm}$, the last
two terms describe their interference.

Distribution (14) contains contributions of
$|\lambda_B^{(f)}|^2$ , ${\rm Re}\lambda_B^{(f)}$
and ${\rm Im}\lambda_B^{(f)}$ multiplied by different functions
of time.  So, at first sight the three quantities can all be
easily extracted if the distribution is experimentally measured
with sufficient accuracy.

However ${\rm Re}\lambda_B^{(f)}$ does not appear
explicitly in distribution (14) if $\Gamma_+$ and $\Gamma_-$
coincide. Thus, a very small expected difference
of $\Gamma_+$ and $\Gamma_-$, contrary to
neutral kaons, may prevent direct measurement of
${\rm Re}\lambda_B^{(f)}$. The situation for
${\rm Im}\lambda_B^{(f)}$ is not so simple as well. In eq.(14)
it is multiplied by $\sin\Delta m_Bt$, which sign is
still unknown since only $|\Delta m_B|$ has been measured.

Therefore, distribution (14) suggests straightforward
measurement for $|\lambda_B^{(f)}|$ and
$|{\rm Im}\lambda_B^{(f)}|$. The sign of
${\rm Im}\lambda_B^{(f)}$ can be measured only in respect
to the sign of $\Delta m_B$.

Surely, even if ${\rm Re}\lambda_B^{(f)}$ can not be directly
measured one is able to calculate
$|{\rm Re}\lambda_B^{(f)}|$ from $|\lambda_B^{(f)}|$
and $|{\rm Im}\lambda_B^{(f)}|$. After that the only
unknown pieces of information on $\lambda_B^{(f)}$ are
the signs of ${\rm Re}\lambda_B^{(f)}$ and
${\rm Im}\lambda_B^{(f)}$.
Let us discuss them in more details.

Note, first of all, that definition (1) can not be used for
unambiguous determination of the states $B_\pm$. Indeed, rephasing
may even interchange the two expressions. So we need some physical
identification for the states. For neutral kaons it was easily done
due to large difference of lifetimes of two neutral kaon states
(i.e., of $K_L$ and $K_S$). But separation of two states by their
lifetimes does not show by itself which of the states $K_S$ and
$K_L$ (or $B_S$ and $B_L$) corresponds to, e.g., $K_+$ (or $B_+$
respectively) in the sense of eq.(1). For this purpose one should
accurately study particular decays (for the kaon case they are pion
decays).

To identify the states $B_\pm$ let us first introduce amplitudes
$a_\pm^{(f)}$ for decays $B_\pm\rightarrow f$. Then we may
rewrite eq.(5) as \begin{equation} \lambda_B^{(f)}\ =\
\frac{a_+^{(f)}-a_-^{(f)}}{a_+^{(f)}+a_-^{(f)}}\,, \end{equation} and \begin{equation}
{\rm
Re}\,\lambda_B^{(f)}\ =\ \frac{|a_+^{(f)}|^2-|a_-^{(f)}|^2}{|a_+^{(f)}+
a_-^{(f)}|^2}\ ;\qquad {\rm Im}\,\lambda_B^{(f)}\ =\ 2\frac{{\rm
Im}(a_+^{(f)} a_-^{(f)*})}{|a_+^{(f)}+a_-^{(f)}|^2}\ .
\end{equation}
Consider, for definiteness, a $CP-$even final state $f+$. If $CP$
were conserved it would be natural to define $B_\pm$ as the states
with $CP-$parity $\pm 1$. Then $a_-^{(f+)}=0$ and $\lambda_B^{(f+)}
=+1$ (for $CP-$odd states $f-$ we would have $a_+^{(f-)}=0$ and
$\lambda_B^{(f-)}=-1$).

In the $CP-$violation case one can not use the $CP-$parity
to identify the states $B_\pm$. But assuming smallness
of the violation we can define $B_+(B_-)$ as being
approximately $CP-$even ($CP-$odd).
It means, by definition, that \begin{equation} |a_+^{(f+)}|\ >\
|a_-^{(f+)}|\,,\qquad|a_-^{(f-)}|\ >\ |a_+^{(f-)}|\ .  \end{equation}
Surely, such a case of approximate $CP-$conservation
leads to the same sign$({\rm Re}\,\lambda_B^{(f)})$ as in
the exact $CP-$conservation case.

Now, without any preliminary assumptions,
we can choose one particular final state $f$ (with a
definite $CP-$parity) and identify states $B_\pm$ by the
corresponding inequality (17). In other words, we ascribe
the $CP-$parity of the particular final state $f$ to be the
approximate $CP-$parity for that of two states $B_\pm$ which
has larger partial width for the decay to $f$.

If the $CP-$violation is small indeed then inequalities
(17) for all other $f$'s are  satisfied as well. However,
if the $CP-$violation is really intrinsically large then
after fixing the states $B_+$ and $B_-$ some of
inequalities (17) might degenerate to equalities
or even reverse their signs (in other words, various
decay channels  might ascribe different values of
approximate $CP-$parity to the same particular state
of the pair $B_\pm$ ).  The latter case can be tested
experimentally by comparing signs of various ${\rm
Re}\lambda_B^{(f)}$ determined from time dependencies (14) for
various final states. It is possible only if the experiment is
exact enough to discriminate $\Gamma_+$ and $\Gamma_-$. Surely,
such an experiment would also allow one to identify two neutral
$B-$meson states by their lifetimes as $B_L$ and $B_S$, just
similar to neutral kaons.

In the absence of such possibility we assume that all
inequalities (17) are correct simultaneously and, thus,
sign$({\rm Re}\,\lambda_B^{(f)})$ is known for any state $f$
being the same as if $CP$ were conserved. After that to make
$\lambda_B^{(f)}$ completely measured one needs to find
sign$({\rm Im}\,\lambda_B^{(f)})$ as well.

Note, first of all, that this sign may be definite only for a
definite choice of the states $B_\pm$. Indeed, according to
eq.(16) their interchange reverses
sign$({\rm Im}\,\lambda_B^{(f)})$. But even if we identified
the states in one way or another we can not fix the sign
by some convention similar to that suggested above for the sign
of ${\rm Re}\lambda_B^{(f)}$. The reason is that in the limit
of $CP-$conservation ${\rm Re}\lambda_B^{(f)}$ tends to the
definite finite limit, while ${\rm Im}\lambda_B^{(f)}$
vanishes. As a result, the sign of ${\rm Re}\lambda_B^{(f)}$
is "kinematic" at not very large $CP-$violation, while the sign
of ${\rm Im}\lambda_B^{(f)}$ is "dynamic" at any degree of the
violation.

Thus, we see that the complete measurement of parameters
$\lambda_B$ for decays (4) requires to determine
sign$({\rm Im}\,\lambda_B)$ from experiment, i.e. from
the corresponding distribution (14). However, such distributions
can only relate signs of ${\rm Im}\lambda_B^{(f)}$
and $\Delta m_B$, but cannot measure them separately.
Therefore, the complete measurement of parameters of direct
$CP-$violation is possible only if one knows sign($\Delta m_B$).

The situation is the same for neutral kaons where
sign(arg$\eta_K$) can be measured only in respect to
sign$(\Delta m_K)$. Essential difference between kaons
and $B-$mesons is much longer lifetime of kaons (even for
$K_S$) which gave possibility to measure
sign$(\Delta m_K)$ in complicated regeneration experiments.
Similar experiments for $B-$mesons are impossible,
and experiments on decays (4) or flavor-tagged decays
(including semileptonic ones) are insensitive to
the absolute sign of $\Delta m_B$ (just as corresponding
decays of neutral kaons).

Thus, neutral $B-$meson decays (4) are able
to demonstrate manifestations of direct $CP-$violation.
But they can provide the complete measurement
for the corresponding $CP-$violating parameters only
if sign$(\Delta m_B)$ is known from some different
experiments.

\section{Measurability of the sign of $\Delta m_B$}

In the preceding Section we discussed only parameters
$\lambda_B$ for decays (4). Parameters $\lambda_{BK}$
for decays (12) studied in \cite{6}  have
similar properties. In particular, signs of various
${\rm Re}\lambda_{BK}$ may be used for identifying states
$B_\pm$ and testing intrinsic smallness of $CP-$violation
by inequalities similar to (17). On the other side,
signs of ${\rm Im}\lambda_{BK}$ can not be fixed by
any convention for the choice of states and should
be determined from experiment. For more detailed
discussion on these parameters see \cite{6}.

Time distributions in decays (12) are more complicated
than distributions (14) in decays (4). They were also
studied in \cite{6}. Here we will not describe them in detail
but summarize two essential points:
\begin{itemize}
\item  The neutral kaon produced in the decay (12)
can be observed only after its own decay by the decay products.
As a result, coherence of $B^0(\overline B^0)$ and
$K^0(\overline K^0)$ evolutions relates the primary decay
(of neutral $B$) and the secondary one (of neutral $K$)
to each other. Distribution in primary $t_1$ and secondary
$t_2$ lifetimes appears, generally, non-factorizable
and depends on the secondary decay mode.
\end{itemize}

E.g., distribution in $t_1$ at $t_2\rightarrow 0$ for kaon
semileptonic decays has the same form as for direct semileptonic
decays of neutral $B-$mesons (though with different normalization).
Two-\-pion kaon decay in the same limit $t_2\rightarrow 0$
produces $t_1-$dependence of the form (14) with substitution
\begin{equation}
\lambda_B\ \rightarrow\ \lambda_{BK}\lambda_K^{\pi\pi}\ ,\qquad
\lambda_K^{\pi\pi}\ =\ \frac{1-\eta_K^{\pi\pi}}{1+\eta_K^{\pi\pi}}\ .
\end{equation}

The opposite extreme case $t_2\rightarrow\infty$ restores
factorization since only $K_L$ survives in the limit. The
corresponding $t_1-$distribution, independent of kaon decay
modes, is given by eq.(14) with
\begin{equation}
\lambda_B\ \rightarrow\  -\ \lambda_{BK}\ .
\end{equation}

\begin{itemize}
\item What is most interesting for purposes of the present paper,
interference of $K_L$ and $K_S$ in the intermediate region of
$t_2$ together with interference of $B_+$ and $B_-$ produces
time distributions sensitive to the sign of $\Delta m_B$
relative to known signs of kaon parameters. This sensitivity
survives even after integration over $t_1$.
\end{itemize}

Therefore, decays (12) allow one not only to search for
$CP-$violation but also to determine experimentally
sign$(\Delta m_B)$ and, thus, provide a necessary basis for
the complete measurement of all parameters of the direct
$CP-$violation in neutral $B-$meson decays. It can be done in
various ways. For example, one can fix both $t_1$ and $t_2$
lifetimes and investigate their correlations in observed time
distributions.  Alternatively, one may not select definite $t_1$
and study only time distribution of secondary kaon decays
integrated over $t_1$. Corresponding general expressions
for both approaches are given in \cite{6}.

As an illustration let us consider here the sequence of decays
\begin{equation}
B^0(\overline B^0)\rightarrow J/\psi K^0(\overline
K^0)\,,\qquad J/\psi\rightarrow \ell^+\ell^-\,,
\qquad K^0(\overline K^0)\rightarrow\pi^+\pi^-\,,
\end{equation}
which has clear experimental manifestation. Using experimental
branching ratios for $B^0\rightarrow J/\psi K^0$ \cite{7},
$J/\psi\rightarrow e^+e^-$, $\mu^+\mu^-$ \cite{8}, $K_S\rightarrow
\pi^+\pi^-$ \cite{8} and the factor 1/2 for $K^0\rightarrow K_S$ we
find the effective branching ratio for events (20) to be equal
\begin{equation}
({\rm Br})_{eff}^{\pi^+\pi^-}\ \approx\ 0.47\cdot 10^{-4}\ .
\end{equation}
According to \cite{6}, the initial pure $B^0-$state produces the
secondary $\pi^+\pi^-$ yield integrated over $t_1$ with the secondary
decay time distribution
\begin{eqnarray}
W_B^{\pi^+\pi^-}(t_2)&\sim& D\cdot\exp(-\Gamma_St_2)+|\eta|^2E
\exp(-\Gamma_Lt_2) \\
&+& 2{\rm Re}\,[\eta\cdot F\exp(-i\Delta m_Kt_2)]\exp
\left(-\frac{\Gamma_L+\Gamma_S}{2}t_2\right)\ , \nonumber
\end{eqnarray}
where
\begin{eqnarray}
D&=&\frac{1}{1-y_B^2}\left(\frac{1+|\lambda|^2}{2}-y_B{\rm Re}\,\lambda
\right)+\frac{1}{1+x_B^2}\left(\frac{1-|\lambda|^2}{2}-
x_B{\rm Im}\,\lambda\right)\ ,\nonumber \\
 E\ & = & \ D(-\lambda)\ , \\
F &=& \frac{1}{1-y_B^2}\left(\frac{1-|\lambda|^2}{2}+iy_B{\rm
Im}\,\lambda\right)+
\frac{1}{1+x_B^2}\left(\frac{1+|\lambda|^2}{2}-ix_B{\rm Re}\,\lambda
\right)\ .    \nonumber
\end{eqnarray}
The notations used here are:
\begin{equation}
y_B\ =\ \frac{\Gamma_+-\Gamma_-}{\Gamma_++\Gamma_-}\ ,\qquad x_B\ =\
2\frac{\Delta m_B}{\Gamma_++\Gamma_-}\ =\ 2\frac{m_--m_+}{\Gamma_++
\Gamma_-}\ ,
\end{equation}
$$\lambda\ \equiv\ \lambda_{BK}^{J/\psi}\ ,\qquad
\eta\ \equiv\ \eta_K^{\pi^+\pi^-}\ .$$
Surely, $\Gamma_L$ and $\Gamma_S$ in eq.(22) are widths of
neutral kaons.

Distribution (22) and its coefficients (23) illustrate
similarity and difference between properties of decays (4)
and (12). Similar to distribution (14), coefficients
$D$ and $E$  are sensitive to the relative signs of
$\Delta \Gamma_B\,=\,\Gamma_+-\Gamma_-$ and
${\rm Re}\lambda_{BK}$, of $\Delta m_B$ and
${\rm Im}\lambda_{BK}$. But the coefficient $F$ has
another structure. It may be obtained from
$D$ by interchange of $y_B$ and $ix_B$. As a
result, this coefficient and the corresponding part of
distribution (22), differently from distribution (14),
are sensitive to the relative signs of $\Delta
\Gamma_B$ and ${\rm Im}\lambda_{BK}$, of $\Delta m_B$ and
${\rm Re}\lambda_{BK}$. Therefore, if we identify the states
$B_\pm$ by fixing the sign of ${\rm Re}\lambda_{BK}$ we
can measure three other signs.

Hence, each particular decay (12), in difference with decays
(4), can provide by itself the complete measurement of the
corresponding parameter $\lambda_{BK}$. Of more universal
interest is that any decay (12) can be used for measuring
sign$(\Delta m_B)$, thus providing necessary information
for the complete measurement of any parameter $\lambda_B$
as well.

For such purposes we may neglect here
$CP-$violation in the primary decay (20) and use
$\lambda_{BK}^{J/\psi}=-1$. We also neglect, for simplicity, the
small quantity $|y_B|$. Then \begin{equation} D\ =\ E\ =\ 1\ ,\qquad F\ =\
\cos\alpha_B\cdot e^{i\alpha_B}\,,\qquad \tan\alpha_B\ =\ x_B\,.
\end{equation} Available data [8] give \begin{equation} |x_B|=0.71\pm
0.06\,,\qquad|\alpha_B|=(35\pm 2)^\circ\,,
\qquad\cos\alpha_B=0.815\pm 0.023\,.  \end{equation} The distribution (22)
may be rewritten now as \begin{eqnarray}
W_B^{\pi^+\pi^-}(t_2)&\sim&\exp(-\Gamma_St_2)+|\eta|^2\exp(-\Gamma_Lt_2)
\\
&+&2|\eta|\cdot\cos\alpha_B\cdot\cos(\alpha_B+\varphi-\Delta
m_Kt_2)\cdot\exp\left(-\frac{\Gamma_S+\Gamma_L}{2}t_2\right)\ ,
\nonumber
\end{eqnarray}
where [8] $$\varphi\ =\ \arg\eta\ =\
\varphi_{+-}\ =\ (44.3\pm 0.8)^\circ\ .$$ The value of $|\alpha_B|$ is
large and comparable to $\varphi$. Thus, two possible sings of
$\alpha_B$ (i.e., of $x_B$) produce very different phases of
oscillation in the third term of distribution (27). However, because of
the small factor $|\eta|\approx 2\cdot 10^{-3}$, their discrimination
requires high experimental statistics.

Therefore, $B-$factories look inappropriate to determine the sign of
$\Delta m_B$ from events (20). More promising might be LHC. The
detector LHC-B dedicated for $B-$physics at LHC \cite{9} is expected
to accumulate 55000 events (20) per year ($10^7$ seconds) at
restricted luminosity $1.5\cdot 10^{32}$cm$^{-2}$s$^{-1}$. We have
used Monte Carlo simulation based on PYTHIA to estimate their
statistical meaning for the above task (note that modifications of
the original PYTHIA were necessary to account for the coupled
coherence of $B-$ and $K-$evolutions; more details will be published
elsewhere). Our results show that reliable determination of
sign$(\Delta m_B)$ requires at least an order more events (20),
i.e. about $10^6$ events. This could be achieved if LHC-B were
modified so to work with higher LHC luminosity.

Another way is to use different sequence of decays
\begin{equation}
B^0(\overline B^0)\rightarrow J/\psi K^0(\overline K^0)\,,
\qquad J/\psi\rightarrow\ell^+\ell^-\,,\qquad K^0(\overline K^0)
\rightarrow\pi^\mp\ell^\pm\nu
\end{equation}
with the effective branching ratio (compare to (21))
\begin{equation}
({\rm Br})_{eff}^{\ell^\pm}\ \approx\  0.45\cdot 10^{-4}\,.
\end{equation}
Its secondary decay time distribution, again integrated over $t_1$
and with $\lambda_{BK}^{J/\psi}=-1$, has the form (compare to (27))
\begin{eqnarray}
W_B^\pm(t_2)&\sim&\exp(-\Gamma_St_2)+\exp(-\Gamma_Lt_2)\  \\
&\pm& 2\cos\alpha_B\cdot\cos(\alpha_B-\Delta m_Kt_2)\cdot\exp
\left(-\frac{\Gamma_S+\Gamma_L}{2}t_2\right)\ . \nonumber
\end{eqnarray}
Here $W_B^\pm$ refers to the secondary lepton $\ell^\pm$ in the
kaon decay. Note that all the distributions (22), (27) and (30)
are written for the initially pure $B^0-$state. For the initial
$\overline B^0-$state one should change the sign of the
interference term in (27) and (30) and, additionally, substitute
$\lambda\rightarrow 1/\lambda$ in (22).

At $\alpha_B=0$ the expression (30) coincides with distributions of
kaon semileptonic decays. In difference with (27), it does not
contain small factor $|\eta|$. Nevertheless, some smallness appears
here as well since only a small part of decays, in the $t_2$
interval of order $\tau_S$, demonstrates oscillations while their
main part, with characteristic time $\tau_L\gg\tau_S$, does
not. Experiments on kaon semileptonic decays show [10] that
oscillations are observable only up to $t_2\la 10^{-9}$s.
Comparing to $\tau_L\approx 5\cdot 10^{-8}$s we see that not
more than 1/50 of all events (28) may be used to extract the
oscillating term. This number noticeably exceeds, however, the
smallness parameter $|\eta|\approx 2\cdot 10^{-3}$ for events
(20).

Therefore, one may hope that determination of the sign of $\Delta
m_B$ will be really achieved at LHC-B or some other facilities by
studying time distribution of events  (28). More reliable estimation
of necessary statistics requires detailed investigation of trigger
efficiencies for such events in a particular detector.

\section{Summary and conclusions}

Here we briefly summarize results of the above considerations.

There are several kinds of $CP-$violating parameters in decays of
neutral $B-$mesons. One of them, $\delta_B$ (see Eq.(3)), is
universal and related to $B\overline{B}$ mixing. However it can
be measured only if the experiment is sensitive enough to
discriminate $\Gamma_L$ and $\Gamma_S$ for $B$-mesons. But even
in such a case, the conventional CKM-\-mechanism of
$CP-$violation strongly suppresses $\delta_B$ and makes it
hardly measurable.

Another set of parameters corresponds to direct $CP$-violation
in $B$-meson decays (4) to final states having definite
$CP$-parity.  It can be identified with phases of quantities
$\lambda_B$ (see Eq.(5)) for the decays with pure elastic
final-\-state rescattering (one of neutral kaon analogues is the
kaon decay to $2\pi$ with  the definite isospin).

One more set of direct $CP-$violating parameters having no
analogues in neutral kaon decays is generated by decays (12) of
neutral $B-$meson to neutral kaon accompanied by a definite
$CP-$parity system (e.g., $J/\psi$). It can be identified with
phases of various $\lambda_{BK}$ (see eq.(13)).

Both values and signs for the direct $CP-$violating phases
are physically meaningful and worth to measure. For example, in
kaon decays only one sign of  arg$\eta$ leads to agreement of
experimental data with the superweak model of $CP-$violation
\cite{11}.  Moreover, the CKM-\-mechanism with 3 quark
generations should unambiguously relate the signs of
$CP-$violating parameters for neutral $B-$meson and neutral
kaon decays (they are all expressible through only one
$CP-$violating parameter of the CKM-\-matrix). However
we demonstrate that any decay (4) by itself can not provide
measurement of the sign of the corresponding $CP-$violating
phase. To achieve the complete measurement of $CP-$violating
parameters one should separately find the sign of $\Delta m_B$.

Therefore, sign$(\Delta m_B)$ appears to be a universal element
which knowledge is very important for studies of $CP-$violation
in neutral $B-$meson decays. We suggest how one can measure the
sign of $\Delta m_B$. This goal may be achieved by extracting
the secondary kaon decay oscillations in the decay sequences
(20) or (28). Monte Carlo simulations show that experimental 
statistics will be insufficient in the near future for events 
(20) with two-\-pion kaon decays. The situation looks more 
promising for events (28) with semileptonic kaon decays.
Corresponding measurements could be done at LHC-B or some other
facilities.

Two of the authors (Ya.I.A. and V.L.R.) thank the International
Science Foundation for support (grants NO~7000 and NO~7300).

\end{document}